\documentclass[
  aps,
  prl,
  twocolumn,
  superscriptaddress,
  nofootinbib,
  floatfix
]{revtex4-2}

\usepackage{amsmath,amssymb,bm}
\usepackage{graphicx}
\usepackage{xcolor}
\usepackage{hyperref}
\usepackage{microtype}

\usepackage{orcidlink}

\hypersetup{
  pdftitle={The Hawking Axion Background},
  pdfauthor={Abdelaziz Hussein, Gordan Krnjaic, Sarunas Verner},
  colorlinks=true,
  citecolor=blue,
  linkcolor=blue,
  urlcolor=blue
}

\newcommand{\mpl}{M_{\rm Pl}}

\newcommand{\ma}{m_a}
\newcommand{\gagg}{g_{a\gamma\gamma}}
\newcommand{\TBH}{T_{\rm BH}}

\newcommand{\supp}{Supplemental Material}
\newcommand{\result}[1]{%
  \ifmmode
    {\color{magenta}\mathbf{[}#1\mathbf{]}}%
  \else
    \textcolor{magenta}{\textbf{[#1]}}%
  \fi
}
\newcommand{\figplaceholder}[2]{%
  \IfFileExists{#1}{%
    \includegraphics[width=\linewidth]{#1}%
  }{%
    \fbox{\parbox[c][#2][c]{0.94\linewidth}{%
      \centering
      \textbf{Figure placeholder}\\[3pt]
      \texttt{\detokenize{#1}}
    }}%
  }%
}

\begin{document}

\hfill{\small FERMILAB-PUB-26-0782-T}

\title{The Cosmology of Hawking Axions}

\author{Abdelaziz Hussein \orcidlink{0000-0002-5560-8668}}
\affiliation{Department of Physics and Kavli Institute for Astrophysics and Space Research,
Massachusetts Institute of Technology, 77 Massachusetts Avenue, Cambridge, MA 02139, USA}

\author{Gordan Krnjaic \orcidlink{0000-0001-7420-9577}}
\affiliation{Theoretical Physics Department, Fermi National Accelerator Laboratory, Batavia, IL 60510, USA}
\affiliation{Department of Astronomy and Astrophysics, University of Chicago, Chicago, IL 60637, USA}
\affiliation{Kavli Institute for Cosmological Physics, University of Chicago, Chicago, IL 60637, USA}

\author{Sarunas Verner \orcidlink{0000-0003-4870-0826}}
\affiliation{Kavli Institute for Cosmological Physics, University of Chicago, Chicago, IL 60637, USA}

\date{\today}

\begin{abstract}
Since primordial black holes (PBH) are always non-relativistic, their cosmological energy fraction grows relative to that of radiation. Thus, for even a tiny initial PBH abundance, the universe generically enters an epoch of early matter domination, which ends when the PBH evaporate to restore the hot big bang. We consider the possibility that a PBH dominated era yields a population of boosted axion-like particles (ALPs) produced from Hawking evaporation. Eventually these ``Hawking axions" become non-relativistic and decay to photons, which can affect various cosmological and astrophysical observables. Here we place new constraints on these particles through their impact on big bang nucleosynthesis, cosmic microwave background, large-scale structure, and X-ray observables. Furthermore, if Hawking axions remain relativistic through recombination, then depending on the progenitor PBH mass, they contribute $\Delta N_{\rm eff}\simeq0.05$--$0.11$, which may be observable with future CMB probes.
\end{abstract}
\maketitle

Primordial black hole (PBH) domination is a compelling and highly-predictive cosmological scenario. Since PBHs always redshift like non-relativistic matter, their energy fraction grows during radiation domination, so the early universe generically becomes PBH dominated unless their initial abundance is extremely fine-tuned. This epoch ends when the PBH evaporate and their Hawking radiation thermalizes to restore radiation domination; nearly all predictions that result from this scenario depend only on the initial PBH mass. For initial masses $M \lesssim 10^9$ g, evaporation occurs before big bang nucleosynthesis (BBN), so this cosmology is largely unconstrained.  

By the equivalence principle, Hawking evaporation produces all known {\it and unknown} particle species. Since Standard Model (SM) particles have appreciable interactions, they promptly thermalize after PBH evaporation, so information about their origins is erased. However, if a new feebly-coupled species were to be produced, it might avoid thermalization with the SM plasma, and thereby retain information from PBH domination as a ``Hawking relic" \cite{Hooper:2019gtx,Baldes:2020nuv,Shallue:2024hqe}. Depending on their mass, Hawking relics can be dark matter candidates \cite{Lennon:2017tqq,Hooper:2019gtx}, increase the effective number of relativistic species \cite{Hooper:2019gtx}, or suppress the growth of structure, in analogy with SM neutrinos \cite{Shallue:2024hqe}. If Hawking relics couple to photons, light ones can convert into X-rays in cosmic magnetic fields, contributing to the cosmic X-ray background (CXB), while heavier ones decay into photon pairs and can alter ionization~\cite{Schiavone:2021imu}.

\begin{figure}[tbp]
 \centering
 \includegraphics[width=0.9\linewidth]{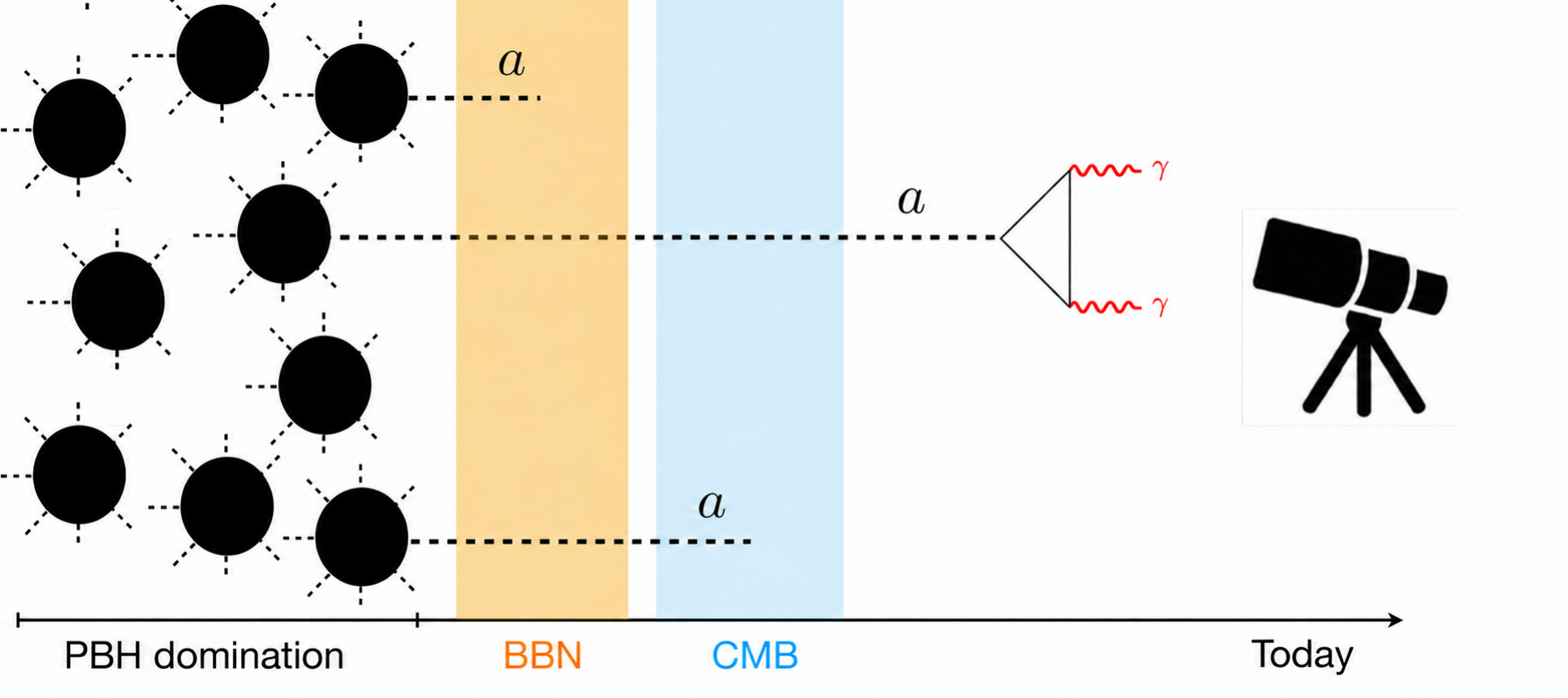}
 \caption{ Schematic illustration of the cosmology studied here. An early epoch of PBH domination ends when Hawking evaporation 
 reheats the universe and SM particles thermalize. If a feebly coupled ALP ($a$) is produced in the evaporation, it can avoid thermalization
 and decay through its photon coupling at late times, yielding cosmological and astrophysical signatures.}
 \label{fig:cartoon}
\end{figure}

In this paper, we systematically analyze the electromagnetic decays of Hawking relic that decay to SM particles and thereby affect cosmological and astrophysical observables (see Fig.~\ref{fig:cartoon}). 
Following Ref.~\cite{Schiavone:2021imu}, without loss of generality, we focus on so-called ``Hawking axions": axion-like particles (ALPs), which are produced via Hawking radiation during PBH domination and eventually decay into photon pairs through the interaction
 \begin{equation}
  \mathcal L\supset
  -\frac{\gagg}{4}aF_{\mu\nu}\widetilde F^{\mu\nu},
  \qquad
  \Gamma_a=\frac{\gagg^2\ma^3}{64\pi},
  \label{eq:interaction}
\end{equation}
where $a$ is the ALP field of mass $\ma$, $\gagg$ is its photon coupling, and $F_{\mu\nu}$ is the electromagnetic field strength and $\tilde{F}^{\mu \nu}$ its dual, and $\Gamma_a$ is the $a\to\gamma\gamma$ decay width in the ALP rest frame. Here we calculate the cosmological ALP abundance produced via Hawking radiation and track its impact on BBN, CMB,  structure formation, and X-ray observables. Our analysis significantly extends the results of Ref.~\cite{Schiavone:2021imu} to include a range of cosmological bounds.

\begin{figure*}[t!]
 \centering
\includegraphics[width=0.495\linewidth]{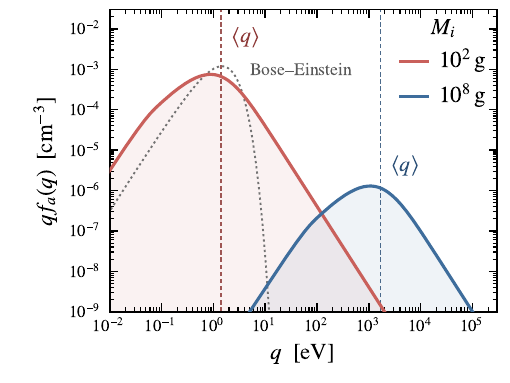}
\includegraphics[width=0.495\linewidth]{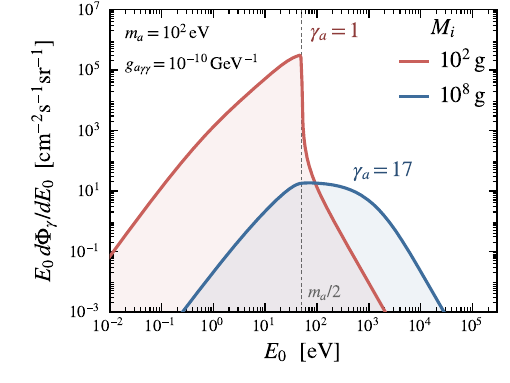}
\caption{{\bf Left:} Comoving momentum distribution $q\,f_a(q)$ for $M_i=10^{2}$ and $10^{8}\,$g,
    evaluated after evaporation with $a_0=1$, so that $q$ is the physical momentum today. Because the PBH heats up as it evaporates, the produced spectra are nonthermal and harder than any single blackbody, with mean momenta $\langle q\rangle\simeq1.4\,$eV and $1.7\,$keV (dashed lines). The dotted curve is a Bose--Einstein spectrum with the same number density and mean momentum as the light case. The mean lies above the peak of $q f_a$ because the final stage of evaporation populates a hard $f_a\propto q^{-3}$ tail. The light case is $\sim\!6\times 10^{2}$ times more abundant and correspondingly softer, $\widehat n_a\propto M_i^{-1/2}$ and $\langle q\rangle\propto M_i^{1/2}$, which accounts for the difference between the constraints in Figs.~\ref{fig:constraints-heavy} and \ref{fig:constraints-light}. {\bf Right:} Present-day photon spectra for $m_a=10^{2}\,$eV and $g_{a\gamma\gamma}=10^{-10}\,{\rm GeV}^{-1}$. The cold light population gives a line ending at $m_a/2$, while the relativistic heavy population gives a broad spectrum.}
 \label{fig:distribution}
\end{figure*}

\vspace{1em}

\textbf{Background Cosmology.}
We consider a monochromatic population of Schwarzschild PBHs with initial mass $M_{i}$, which dominate the energy density and evaporate before BBN. This limits the PBH mass range to $1 {\rm g}\lesssim M_i \lesssim5\times10^{8}{\rm g}$, where the lower limit follows from requiring the PBHs to form after inflation~\cite{Planck:2018jri} and the upper limit ensures that they have evaporated before BBN~\cite{Carr:2020gox}. If we sum over SM species, the Hawking mass-loss rate scales as
\begin{equation}
\label{eq:massloss-main}
\frac{dM}{dt} = - \frac{{\cal G} g_{\star,H}(T_{\rm BH})  }{30720 \pi G^2 M^2}~~,~~  \TBH=\frac{1}{8\pi G M},
\end{equation}
where $T_{\rm BH}$ is the Hawking temperature, $G$ is the Newton constant,
${\cal G}\approx3.8$ is the graybody factor, and $g_{\star,H}$ is the spin-weighted
number of relativistic degrees of freedom, with $g_{\star,H}\simeq108$ for the SM~\cite{Hooper:2019gtx}.

The background evolves according to
\begin{align}
 \dot\rho_{\rm BH}+3H\rho_{\rm BH}
 =\frac{\dot M}{M}\rho_{\rm BH}~,~
 \dot\rho_R+4H\rho_R
 =-\frac{\dot M}{M}\rho_{\rm BH},
 \label{eq:background-main}
\end{align}
where $\rho_{R} = \pi^2 g_\star T^4/30$ is the SM radiation density, $g_\star$ is the effective number of relativistic species,  $\rho_{\rm BH}$ is the PBH density, and $H^2\propto (\rho_{\rm BH}+\rho_R)$ is the Hubble rate. Since PBH domination is an attractor solution for even a tiny initial abundance~\cite{Hooper:2019gtx}, we assume the initial conditions: $\rho_{R,i}= 0$  and $\rho_{{\rm BH},i} = 3 H_i^2/ (8\pi G)$, noting that the system is insensitive to the latter condition for sufficiently large values of $H_i$. When the PBH evaporate, the universe is reheated to a temperature \cite{Hooper:2019gtx}
\begin{equation}
\hspace{-0.2cm}
T_{\rm RH} \approx 50 \, {\rm MeV} \! \left(     \frac{10^8 \rm \, g}{M_i } \right)^{3/2} 
\!
 \left(     \frac{14}{g_{\star}} \right)^{1/4}
 \!
 \left(     \frac{g_{\star, H} }{108} \right)^{1/2}
\! ,
 \end{equation}
where $g_{\star, H} $ and $g_{\star}$ are evaluated at $T_{{\rm BH}, i} = (8\pi GM_i)^{-1}$ and  $T_{\rm RH}$, respectively, and we have used the instantaneous evaporation approximation, though in our main results we solve Eq.~\eqref{eq:background-main} numerically. 

\vspace{1em}

\textbf{Hawking Axion Population.}
For ALP production, the instantaneous Hawking spectrum is
\begin{equation}
 \frac{d^2N_a}{dE_a\,dt}
 =\frac{1}{2\pi}
 \frac{\Gamma_0(E_a,M)}
      {\exp[E_a/\TBH(M)]-1},
 ~~
 \label{eq:hawking-main}
\end{equation}
where $\TBH$ is the Hawking temperature of a black hole with instantaneous mass $M$, $\Gamma_0$ is the dimensionless spin-zero graybody factor, and  $E_a=\sqrt{p^2 + \ma^2}$.
We integrate Eq.~\eqref{eq:hawking-main} from PBH formation\footnote{As long as $t_f \ll t_{\rm ev},$ the integral is insensitive to $t_f$, which can be taken to $t_f \to 0.$} at $t_f$ until complete evaporation at $t_{\rm ev}$, while evolving Eq.~\eqref{eq:background-main} through reheating. 

Since the Hawking temperature increases as the PBH loses its mass, the resulting spectrum produces a hard power-law tail that is absent in a fixed mass approximation. The result is the comoving ALP distribution $f_a(q)\equiv d\widehat n_a/dq$, with $q=ap$, where $a$ is the scale factor normalized according to $a_0=1$:
\begin{equation}
 f_a(q)\approx \widehat n_{\rm BH}
 \int_{0}^{t_{\rm ev}}\frac{dt}{a(t)}
 \frac{p}{E_a}
 \left.\frac{d^2N_a}{dE_a\,dt}\right|_{p=q/a(t)},
 \label{eq:phase-space-main}
\end{equation}
where $1/a$ and $p/E_a$ are the comoving momentum and massive particle Jacobians and we have taken the $t_f \ll t_{\rm ev}$ limit. The comoving PBH density $\widehat n_{\rm BH}=a^3n_{\rm BH}$ is fixed by the same background and reheating solution, so $f_a$ already contains both the PBH abundance and the ALP yield per PBH, so no additional normalization is needed in the photon flux. We then use numerical scalar graybody factors throughout and keep the geometric optics approximation $\Gamma_0\to27G^2M^2E_a^2$ only as an analytic cross-check~\cite{Arbey:2019mbc,Arbey:2021mbl}. Details of the PBH background, the attractor solution, and the phase space calculation are given in the~\supp.

\begin{figure*}[tbp]
 \centering
 \includegraphics[width=0.9\linewidth]{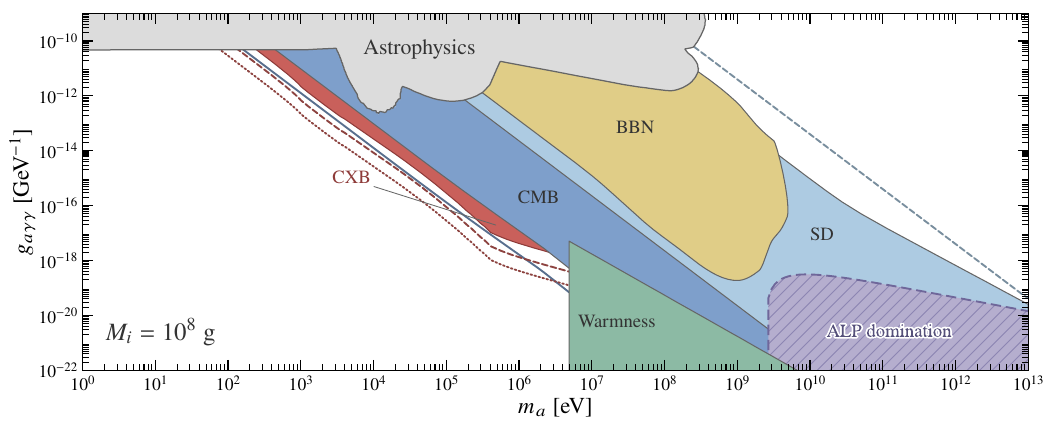}
 \caption{Constraints on decaying Hawking ALPs for
 $M_i=10^{8}\,{\rm g}$. All filled regions are derived from the same comoving ALP  distribution: the direct CXB overshoot bound (red), defined by Eq.~\eqref{eq:overshoot-main}, Planck CMB anisotropies (dark blue), BBN photodisintegration (yellow), FIRAS $\mu$ and $y$ spectral distortions (light blue)~\cite{Fixsen:1996nj,Chluba:2011hw}, and the recast structure formation bound on stable relics (green), applied only where decays are negligible~\cite{Shallue:2024hqe}. In the hatched purple region, the ALPs come to dominate the energy density and the fixed background calculation is no longer consistent and this region is excluded. Gray shading shows existing laboratory and astrophysical limits, which do not depend on $f_a$~\cite{AxionLimits,Ayala:2014pea,Dolan:2022kul,Payez:2014xsa,CAST:2017uph}. Lines show projected sensitivities. The dashed and dotted red curves give the direct photon reach if the astrophysical CXB can be modeled and subtracted to $10\%$ and $1\%$ accuracy, respectively. The solid blue curve is the Simons Observatory projection for the CMB anisotropy recast~\cite{SimonsObservatory:2018koc}, and the dashed blue line is a PIXIE-like spectral distortion forecast~\cite{Kogut:2011xw}. 
 }
 \label{fig:constraints-heavy}
\end{figure*}

\vspace{1em}

{\textbf{Propagation and photon signal.}}
After evaporation ends at the reheating redshift $z_{\rm RH}$, the ALPs free stream with fixed comoving momentum $q$, so that $p(q,z)=q(1+z)$ and $E_a(q,z)=\sqrt{ p(q,z)^2 +\ma^2 }$. In the frame 
of a comoving observer, the decay rate is $\Gamma_a/\gamma_a(q,z)$, with $\gamma_a=E_a/\ma$, and the probability that an ALP survives from $z_{\rm RH}$ to $z$ is 
\begin{equation}
 S_a(q,z)=\exp\!\left[-\!\int_z^{z_{\rm RH}}\!\!
 \frac{dz'}{(1+z')H(z')}\,
 \frac{\Gamma_a}{\gamma_a(q,z')}\right] \,.
 \label{eq:survival-main}
\end{equation}
As the momentum redshifts, the Lorentz factor decreases and the decay rate in this frame grows. Since the decay in the ALP rest frame is isotropic, a single decay produces a box spectrum in the comoving frame,
\begin{equation}
 K_\gamma(E_\gamma;q,z)
 =\frac{2}{p}\,
 \Theta(E_\gamma-E_-)\Theta(E_+-E_\gamma),
 \label{eq:box-main}
\end{equation}
where $E_\pm=\frac{1}{2}(E_a\pm p)$ and $K_\gamma \to 2\delta\left(E_\gamma-\frac{1}{2}\ma\right)$ in the nonrelativistic limit. Combining Eqs.~\eqref{eq:survival-main} and \eqref{eq:box-main} with the distribution of Eq.~\eqref{eq:phase-space-main} gives the comoving photon emissivity, the number of photons produced per unit time, energy, and comoving volume,
\begin{equation}
 \widehat Q_\gamma(E_\gamma,z)=
\Gamma_a  \int_0^\infty \frac{ dq \, f_a(q) }{\gamma_a(q,z)}  \,\,S_a(q,z) \,
 K_\gamma(E_\gamma;q,z) .
 \label{eq:emissivity-main}
\end{equation}
We describe photon attenuation between emission at redshift $z$ and observation today by $e^{-\tau_\gamma(E_0,z)}$, with the optical depth evaluated for a photon observed at energy $E_0$. The present-day differential photon flux is then
\begin{equation}
 \frac{d\Phi_\gamma}{dE_0}
 =\frac{1}{4\pi}\int_0^{z_{\rm RH}}\frac{dz}{H(z)}\,
 e^{-\tau_\gamma(E_0,z)}\,
 \widehat Q_\gamma[E_0(1+z),z] \, ,
 \label{eq:flux-main}
\end{equation}
with no additional factor of $(1+z)$, since the factor $(1+z)^{-1}$ in $|dt/dz|$ cancels the Jacobian $dE_\gamma/dE_0=1+z$. The opacity model and the numerical checks are given in the \supp.

We show in Fig.~\ref{fig:distribution} the ALP momentum distributions for the two cases discussed
below (left) and the resulting photon spectra (right).
If most ALPs decay while still relativistic, the observed spectrum retains the Hawking momentum scale and is much broader than a delta function. If instead the population becomes nonrelativistic before decaying, the emitted spectrum narrows around $\ma/2$, and the observed photon energies are set by the decay redshift. Describing the transition between these regimes requires the full momentum distribution and its momentum-dependent decay history.

\vspace{1em}

{\textbf{Observational constraints.}}
Laboratory and astrophysical limits on $\gagg$ apply directly to our scenario and do not depend on the cosmological Hawking population~\cite{AxionLimits,Ayala:2014pea,Dolan:2022kul,Payez:2014xsa,
CAST:2017uph}. For all other bounds, we use $f_a(q)$ from Eq.~\eqref{eq:phase-space-main} and $S_a(q,z)$ from Eq.~\eqref{eq:survival-main} to compute the surviving ALP energy density,
\begin{equation}
 \rho_a(z)=(1+z)^3\int_0^\infty dq\,f_a S_a E_a\, ,
 \label{eq:inputs-main1}
\end{equation}
and the energy injected per unit time and volume, obtained by summing the photon energies,
\begin{equation}
 \dot\rho_{\rm inj}(z)=(1+z)^3\!\int_0^\infty dE_\gamma\,E_\gamma\,\widehat Q_\gamma(E_\gamma,z) .
 \label{eq:inputs-main2}
\end{equation}
BBN and the CXB depend on the full photon spectrum, the CMB constraints mainly on the injected
energy, and ALP domination on $\rho_a$:

\begin{figure*}[tbp]
 \centering
 \includegraphics[width=0.9\linewidth]{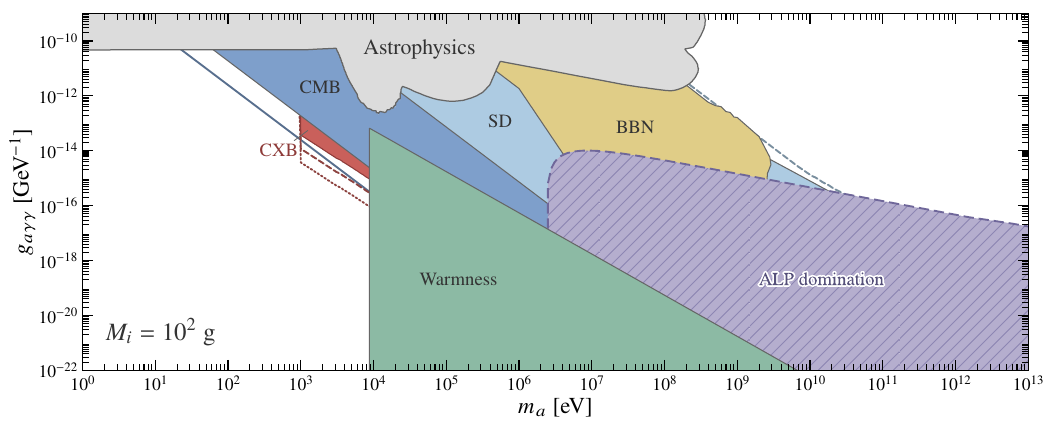}
 \caption{Same as Fig.~\ref{fig:constraints-heavy}, for
 $M_i=10^{2}\,$g. The lighter PBHs evaporate earlier, and leave a softer relic population with a number density $\sim6\times10^{2}$ times larger. Secondary ALP domination (hatched purple) begins at $\ma\gtrsim2.5\,\mathrm{MeV}$ and covers the long-lived region, including the CXB band at these masses. The structure formation exclusion (green) begins at the stable relic limit $\ma\simeq10\,\mathrm{keV}$~\cite{Shallue:2024hqe}. The direct CXB exclusion shrinks to a narrow red sliver near $\ma\sim1\,\mathrm{keV}$, where late decays of the abundant
 cold population overshoot the soft X-ray background. The BBN and spectral distortion constraints remain important at high masses. Line styles are as in
 Fig.~\ref{fig:constraints-heavy}. The PIXIE and Simons Observatory
 projections extend the reach beyond the current
 spectral distortion and CMB anisotropy bounds, respectively.}
 \label{fig:constraints-light}
\end{figure*}

\begin{itemize}

\item{\bf ALP Domination:} Heavy and long-lived ALPs redshift like matter and can come to dominate the energy density. If they survive until today they exceed the observed matter density, and if they decay after BBN they inject more energy into photons than the entire radiation background, so we exclude every point where $\rho_a$ exceeds the background density at any redshift.

\item{\bf BBN:}
Decay photons injected after BBN ($z\lesssim2\times10^{7}$) can break up light nuclei and change their abundances. We pass $\widehat Q_\gamma$ to the electromagnetic cascade code \textsc{ACROPOLIS}~\cite{Hufnagel:2018bjp,Depta:2020zbh,Depta:2020mhj}, which follows the photon cascade and the resulting abundances, and exclude points where D/H or $Y_p$ is outside of $95\%$ allowed range or $^3$He/D exceeds its upper bound.

\item{\bf CMB Spectral Distortions:} Energy injected between $z\simeq2\times10^{6}$ and recombination
distorts the CMB blackbody spectrum. This effect is quantified with spectral distortion parameters 
\begin{equation}
\hspace{1cm}\mu\simeq1.4\!\int\!dt\,{\cal J}_\mu\frac{\dot\rho_{\rm inj}}{\rho_\gamma},\qquad
 y\simeq\frac14\!\int\!dt\,{\cal J}_y\frac{\dot\rho_{\rm inj}}{\rho_\gamma},
 \label{eq:mu-y-main}
\end{equation}
with visibility functions taken from Refs. ${\cal J}_{\mu,y}$~\cite{Chluba:2011hw,Chluba:2013wsa}. We exclude ALP parameter space  in which these distortions exceed the FIRAS $95\%$ exclusion limits, corresponding to 
$|\mu|>9\times10^{-5}$ or $|y|>1.5\times10^{-5}$~\cite{Fixsen:1996nj}.

\item{\bf CMB Temperature Anisotropies:} Energy injected near recombination alters the ionization history. Planck constrains dark matter decaying into two photons through the product of its abundance and decay rate~\cite{Slatyer:2016qyl,Poulin:2016anj,Cang:2020exa}. We apply this limit with the ALP mass density in place of the dark matter density, average it over the box spectrum weighted by photon energy, and use it only where the ALPs are nearly cold and long-lived. Following Ref.~\cite{Langhoff:2022bij}, we evaluate the box average and the ALP survival at a single redshift, $z_*=300$, and assume that the effect on the CMB scales linearly with the injected energy, so the published limit can be rescaled without a new recombination and likelihood calculation for each point (see \supp).

\item{\bf CXB:} Photons produced from Hawking axion decays can contribute to the cosmic  X-ray background (CXB). We compare their energy-weighted intensity ${\cal I}(E_0)=E_0^2\,d\Phi_\gamma/dE_0$ from Eq.~\eqref{eq:flux-main},
which includes attenuation, with a reference CXB spectrum ${\cal B}_{\rm CXB}(E_0)$ built from the total CXB fits of Refs.~\cite{Cappelluti_2017,Moretti:2008hs}, and exclude any point where
\begin{equation}
 {\cal I}(E_0)>{\cal B}_{\rm CXB}(E_0),
 \label{eq:overshoot-main}
\end{equation}
for $0.5 \, {\rm keV}\leq E_0\leq200 \, {\rm keV}$, that is, where the ALP photons alone would exceed the measured X-ray background. Since this lets the ALPs account for the entire background without subtracting any astrophysical sources, the bound is conservative, although it is not a limit at a fixed confidence level. Two further choices make the bound conservative: the reference spectrum lies at or above most of the CXB measurements from Ref.~\cite{Ajello:2008xb}, so the ALPs must exceed a larger background, and we include only the extragalactic signal, neglecting photons from ALPs decaying in the Galactic halo. In the absence of attenuation, integrating Eq.~\eqref{eq:flux-main} over all observed energies gives
\begin{equation}
\hspace{1cm}
\int dE_0\,\frac{d\Phi_\gamma}{dE_0}
 =\frac{1}{2\pi}\int dq\,f_a(q)\,[1-S_a(q,0)],
 \label{eq:number-check-main}
\end{equation}
two photons for each ALP that has decayed by today, which we use to check the numerical flux.

\item{\bf Structure formation:} 
 Long-lived Hawking relics can suppress structure formation in the late universe, in analogy with the familiar effect of massive neutrinos in standard cosmology. In Figs. \ref{fig:constraints-heavy} and \ref{fig:constraints-light} we apply the constraints on this scenario calculated in  Ref.~\cite{Shallue:2024hqe}.
\end{itemize}

Alongside the current bounds shown in Figs. \ref{fig:constraints-heavy}, we also include the projected reach of a PIXIE-like spectral distortion mission~\cite{Kogut:2011xw} and the Simons Observatory sensitivity to CMB anisotropies~\cite{SimonsObservatory:2018koc}. For the direct photon channel, we show the sensitivity that could be achieved if the astrophysical X-ray background were modeled and subtracted. Each procedure, its domain of validity, and its statistics are specified in the \supp.

We show in Fig.~\ref{fig:constraints-heavy} the heavy  PBH case, $M_i=10^{8}\,$g, for which the direct photon channel is strongest. The CXB exclusion is a narrow band whose lower boundary is set by the lifetime, since at smaller couplings too few ALPs have decayed by today. At larger couplings, the decays occur too early for the photons to reach us, so cosmic opacity sets the upper boundary. As increasing $\gagg$ shifts the decays to higher redshift, the dominant constraint changes from CMB anisotropies to $y$ and $\mu$ distortions and finally to BBN photodisintegration. At MeV masses, the combined exclusion spans roughly seven decades in coupling.

Ref.~\cite{Schiavone:2021imu} focused on ultralight ALPs converting in cosmic magnetic fields and treated decays of ALPs with $10\,{\rm eV}\lesssim\ma\lesssim1\,$MeV more briefly, including the Lorentz-dilated decay rate and boosted two-photon kinematics, with constraints from reionization. Here we compute the ALP distribution with numerical spin-zero graybody factors and the full PBH-dominated background, derive a dedicated CXB bound on decaying Hawking axions, and combine it in Figs.~\ref{fig:constraints-heavy} and~\ref{fig:constraints-light} with BBN, CMB, structure formation, and ALP-domination constraints that extend to $\ma\sim10\,$TeV.

For this case, typical relic momenta today are of order $\mathrm{keV}$, so boosted decays can therefore populate the X-ray band even when $\ma/2$ lies below its lower edge, which extends the CXB exclusion down to $\ma\sim10^{2}\,\mathrm{eV}$. In the same way, photodisintegration likewise depends on the boosted photon energy rather than on $\ma/2$, and the BBN constraint extends below the threshold for decays at rest, $\ma=2E_{\rm th}^{D}\simeq4.4\,\mathrm{MeV}$. At large masses, the spectral distortion wedge extends to $\ma\sim10\,\mathrm{TeV}$, where $\ma/T_{\rm BH}^{\,i}\simeq0.1$ and ALP emission is only mildly suppressed.

In Fig.~\ref{fig:constraints-light} we show the same results for the  light PBH case, $M_i=10^{2}\,\mathrm{g}$. Since the comoving abundance scales approximately as $M_i^{-1/2}$, and the characteristic momentum  scales as $M_i^{1/2}$, at fixed ALP mass, lighter PBHs yield a colder and more abundant relic population. For a stable ALP, $\Omega_ah^2\simeq0.20\,(\ma/{\rm MeV})$ at this benchmark, and even a decayed fraction of $\sim10^{-7}$ is enough to saturate the CXB. As a result, the exclusion is dominated by self-consistency rather than by the photon flux. Over most of the long-lived parameter space the ALPs would dominate the energy density before decaying, and the would-be direct band lies inside this region. Comparing Figs.~\ref{fig:constraints-heavy} and~\ref{fig:constraints-light}, the PBH mass determines which probe dominates in each part of the plane and sets the overall normalization.

\textbf{Discussion.}
The results in Figs.~\ref{fig:constraints-heavy} and~\ref{fig:constraints-light} contain no free abundance parameter. For fixed PBH and ALP masses, the PBH-dominated history determines the initial relic phase space, and $\gagg$ enters only through the decay. Direct photons, energy deposition before and after recombination, structure formation, and consistency of the assumed cosmological history constrain complementary regions of the mass--coupling plane, with a relative importance that depends on $M_i$.

Future measurements can extend these limits. A PIXIE-like mission could push the spectral distortion bounds to smaller couplings for ALP masses up to $\ma\sim10\,\mathrm{TeV}$ and Simons Observatory is projected to improve on the Planck recast. Better modeling and subtraction of the astrophysical CXB would also strengthen the direct photon bound. For long-lived ALPs, the $10\%$ background uncertainty assumed here would probe couplings about $2.5$ times smaller. In the limit $\gagg\to0$, a population that remains relativistic through recombination contributes $\Delta N_{\rm eff}\simeq0.05$--$0.11$ across the PBH mass window, which is an additional target for future CMB measurements~\cite{SimonsObservatory:2018koc}.

Our calculation assumes a monochromatic population of Schwarzschild PBHs, semiclassical evaporation, and PBH domination before evaporation. We use massless ALP graybody tables for $\ma/T_{\rm BH}^{\,i}\lesssim0.1$, require ALP decays during evaporation to be negligible, and neglect plasma-induced reabsorption. The validity conditions and possible extensions are discussed in the Supplemental Material. A joint likelihood using the measured CXB spectrum could improve the direct photon limit, and a full electromagnetic transfer calculation would treat photon propagation, absorption, and energy deposition consistently across the CMB and BBN observables. With the appropriate Hawking spectrum, decay width, and treatment of the daughter particles, the same method applies to other unstable particles produced by Hawking evaporation.

Because the ALP population is produced gravitationally, its abundance does not vanish as $\gagg$ decreases. Diffuse photon measurements can then probe couplings at which production through the ALP interaction itself would be negligible. More generally, these observations provide a way to test unstable Hawking relics and the particle content of the Universe before BBN.

\medskip

\begin{acknowledgments}
{\textit{ \bf Acknowledgments.}} We thank Mainak Mukhopadhyay, whose comments improved this manuscript by an amount we are unable to bound from above. This work made use of the \textit{AxionLimits} repository \cite{AxionLimits} by C. O'Hare. This manuscript has been authored by Fermi Forward Discovery Group, LLC under Contract No. 89243024CSC000002 with the U.S. Department of Energy, Office of Science, Office of High Energy Physics. A.H. acknowledges that this material is based upon work supported by the U.S. Department of Energy, Office of Science, Office of Workforce Development for Teachers and Scientists, Office of Science Graduate Student Research (SCGSR) program. The SCGSR program is administered by the Oak Ridge Institute for Science and Education (ORISE) for the DOE. ORISE is managed by ORAU under contract number DESC0014664. All opinions expressed in this paper are the author’s and do not necessarily reflect the policies and views of DOE, ORAU, or ORISE. Fermilab is operated by Fermi Research Alliance, LLC, under Contract No. DE-AC02-07CH11359 with the US Department of Energy. The work of S.V. was supported by the Kavli Institute for Cosmological Physics at the University of Chicago.
\end{acknowledgments}

\bibliographystyle{apsrev4-2}
\bibliography{references.bib}

@article{Page:1976df,
    author = "Page, Don N.",
    title = "{Particle Emission Rates from a Black Hole: Massless Particles from an Uncharged, Nonrotating Hole}",
    doi = "10.1103/PhysRevD.13.198",
    journal = "Phys. Rev. D",
    volume = "13",
    pages = "198--206",
    year = "1976"
}

@article{Page:1976ki,
    author = "Page, Don N.",
    title = "{Particle Emission Rates from a Black Hole. 2. Massless Particles from a Rotating Hole}",
    doi = "10.1103/PhysRevD.14.3260",
    journal = "Phys. Rev. D",
    volume = "14",
    pages = "3260--3273",
    year = "1976"
}

@article{Hooper:2019gtx,
    author = "Hooper, Dan and Krnjaic, Gordan and McDermott, Samuel D.",
    title = "{Dark Radiation and Superheavy Dark Matter from Black Hole Domination}",
    eprint = "1905.01301",
    archivePrefix = "arXiv",
    primaryClass = "hep-ph",
    reportNumber = "FERMILAB-PUB-19-186-A",
    doi = "10.1007/JHEP08(2019)001",
    journal = "JHEP",
    volume = "08",
    pages = "001",
    year = "2019"
}

@article{MacGibbon:1990zk,
    author = "MacGibbon, J. H. and Webber, B. R.",
    title = "{Quark and gluon jet emission from primordial black holes: The instantaneous spectra}",
    doi = "10.1103/PhysRevD.41.3052",
    journal = "Phys. Rev. D",
    volume = "41",
    pages = "3052--3079",
    year = "1990"
}

@article{Lennon:2017tqq,
    author = "Lennon, Olivier and March-Russell, John and Petrossian-Byrne, Rudin and Tillim, Hannah",
    title = "{Black Hole Genesis of Dark Matter}",
    eprint = "1712.07664",
    archivePrefix = "arXiv",
    primaryClass = "hep-ph",
    doi = "10.1088/1475-7516/2018/04/009",
    journal = "JCAP",
    volume = "04",
    pages = "009",
    year = "2018"
}

@article{deSalas:2015glj,
    author = "de Salas, P. F. and Lattanzi, M. and Mangano, G. and Miele, G. and Pastor, S. and Pisanti, O.",
    title = "{Bounds on very low reheating scenarios after Planck}",
    eprint = "1511.00672",
    archivePrefix = "arXiv",
    primaryClass = "astro-ph.CO",
    reportNumber = "IFIC-15-70",
    doi = "10.1103/PhysRevD.92.123534",
    journal = "Phys. Rev. D",
    volume = "92",
    number = "12",
    pages = "123534",
    year = "2015"
}

@article{Hasegawa:2019jsa,
    author = "Hasegawa, Takuya and Hiroshima, Nagisa and Kohri, Kazunori and Hansen, Rasmus S. L. and Tram, Thomas and Hannestad, Steen",
    title = "{MeV-scale reheating temperature and thermalization of oscillating neutrinos by radiative and hadronic decays of massive particles}",
    eprint = "1908.10189",
    archivePrefix = "arXiv",
    primaryClass = "hep-ph",
    reportNumber = "KEK-TH-2149, KEK-Cosmo-242, RIKEN-iTHEMS-Report-19, IPMU19-0120",
    doi = "10.1088/1475-7516/2019/12/012",
    journal = "JCAP",
    volume = "12",
    pages = "012",
    year = "2019"
}

@article{Arbey:2019mbc,
    author = "Arbey, Alexandre and Auffinger, J{\'e}r{\'e}my",
    title = "{BlackHawk: A public code for calculating the Hawking evaporation spectra of any black hole distribution}",
    eprint = "1905.04268",
    archivePrefix = "arXiv",
    primaryClass = "gr-qc",
    reportNumber = "CERN-TH-2019-067",
    doi = "10.1140/epjc/s10052-019-7161-1",
    journal = "Eur. Phys. J. C",
    volume = "79",
    number = "8",
    pages = "693",
    year = "2019"
}

@article{Arbey:2021mbl,
    author = "Arbey, Alexandre and Auffinger, J{\'e}r{\'e}my",
    title = "{Physics Beyond the Standard Model with BlackHawk v2.0}",
    eprint = "2108.02737",
    archivePrefix = "arXiv",
    primaryClass = "gr-qc",
    reportNumber = "CERN-TH-2021-117",
    doi = "10.1140/epjc/s10052-021-09702-8",
    journal = "Eur. Phys. J. C",
    volume = "81",
    pages = "910",
    year = "2021"
}

@article{Chluba:2011hw,
    author = "Chluba, J. and Sunyaev, R. A.",
    title = "{The evolution of CMB spectral distortions in the early Universe}",
    eprint = "1109.6552",
    archivePrefix = "arXiv",
    primaryClass = "astro-ph.CO",
    doi = "10.1111/j.1365-2966.2011.19786.x",
    journal = "Mon. Not. Roy. Astron. Soc.",
    volume = "419",
    pages = "1294--1314",
    year = "2012"
}

@article{Fixsen:1996nj,
    author = "Fixsen, D. J. and Cheng, E. S. and Gales, J. M. and Mather, John C. and Shafer, R. A. and Wright, E. L.",
    title = "{The Cosmic Microwave Background spectrum from the full COBE FIRAS data set}",
    eprint = "astro-ph/9605054",
    archivePrefix = "arXiv",
    doi = "10.1086/178173",
    journal = "Astrophys. J.",
    volume = "473",
    pages = "576",
    year = "1996"
}

@article{Kogut:2011xw,
    author = "Kogut, A. and others",
    title = "{The Primordial Inflation Explorer (PIXIE): A Nulling Polarimeter for Cosmic Microwave Background Observations}",
    eprint = "1105.2044",
    archivePrefix = "arXiv",
    primaryClass = "astro-ph.CO",
    doi = "10.1088/1475-7516/2011/07/025",
    journal = "JCAP",
    volume = "07",
    pages = "025",
    year = "2011"
}

@article{Cappelluti_2017,
   title={The Chandra COSMOS Legacy Survey: Energy Spectrum of the Cosmic X-Ray Background and Constraints on Undetected Populations},
   volume={837},
   ISSN={1538-4357},
   url={http://dx.doi.org/10.3847/1538-4357/aa5ea4},
   DOI={10.3847/1538-4357/aa5ea4},
   number={1},
   journal={The Astrophysical Journal},
   publisher={American Astronomical Society},
   author={Cappelluti, Nico and Li, Yanxia and Ricarte, Angelo and Agarwal, Bhaskar and Allevato, Viola and Ananna, Tonima Tasnim and Ajello, Marco and Civano, Francesca and Comastri, Andrea and Elvis, Martin and Finoguenov, Alexis and Gilli, Roberto and Hasinger, Günther and Marchesi, Stefano and Natarajan, Priyamvada and Pacucci, Fabio and Treister, E. and Urry, C. Megan},
   year={2017},
   month=feb, pages={19} }

@misc{AxionLimits,
  author       = {Ciaran O'Hare},
  title        = {cajohare/AxionLimits: AxionLimits},
  month        = jul,
  year         = 2020,
  publisher    = {Zenodo},
  version      = {v1.0},
  doi          = {10.5281/zenodo.3932430},
  howpublished = {\url{https://cajohare.github.io/AxionLimits/}}
}

@article{DeLuca:2019buf,
    author = "De Luca, V. and Desjacques, V. and Franciolini, G. and Malhotra, A. and Riotto, A.",
    title = "{The initial spin probability distribution of primordial black holes}",
    eprint = "1903.01179",
    archivePrefix = "arXiv",
    primaryClass = "astro-ph.CO",
    doi = "10.1088/1475-7516/2019/05/018",
    journal = "JCAP",
    volume = "05",
    pages = "018",
    year = "2019"
}

@article{Mirbabayi:2019uph,
    author = "Mirbabayi, Mehrdad and Gruzinov, Andrei and Nore{\~n}a, Jorge",
    title = "{Spin of Primordial Black Holes}",
    eprint = "1901.05963",
    archivePrefix = "arXiv",
    primaryClass = "astro-ph.CO",
    doi = "10.1088/1475-7516/2020/03/017",
    journal = "JCAP",
    volume = "03",
    pages = "017",
    year = "2020"
}

@article{Arbey:2021ysg,
    author = "Arbey, Alexandre and Auffinger, J{\'e}r{\'e}my and Sandick, Pearl and Shams Es Haghi, Barmak and Sinha, Kuver",
    title = "{Precision calculation of dark radiation from spinning primordial black holes and early matter-dominated eras}",
    eprint = "2104.04051",
    archivePrefix = "arXiv",
    primaryClass = "astro-ph.CO",
    reportNumber = "CERN-TH-2021-049",
    doi = "10.1103/PhysRevD.103.123549",
    journal = "Phys. Rev. D",
    volume = "103",
    number = "12",
    pages = "123549",
    year = "2021"
}

@article{Barrow:1990he,
    author = "Barrow, John D. and Copeland, Edmund J. and Kolb, Edward W. and Liddle, Andrew R.",
    title = "{Baryogenesis in Extended Inflation.II. Baryogenesis via Primordial Black Holes}",
    reportNumber = "FERMILAB-PUB-90-098-A, SUSSEX-AST-90-7-1",
    doi = "10.1103/PhysRevD.43.984",
    journal = "Phys. Rev. D",
    volume = "43",
    pages = "984--994",
    year = "1991"
}

@article{Shallue:2024hqe,
    author = "Shallue, Christopher J. and Mu{\~n}oz, Julian B. and Krnjaic, Gordan Z.",
    title = "{Warm Hawking relics from primordial black hole domination}",
    eprint = "2406.08535",
    archivePrefix = "arXiv",
    primaryClass = "astro-ph.CO",
    reportNumber = "FERMILAB-PUB-24-0298-T",
    doi = "10.1088/1475-7516/2025/02/026",
    journal = "JCAP",
    volume = "02",
    pages = "026",
    year = "2025"
}

@article{Cheek:2022mmy,
    author = "Cheek, Andrew and Heurtier, Lucien and Perez-Gonzalez, Yuber F. and Turner, Jessica",
    title = "{Evaporation of primordial black holes in the early Universe: Mass and spin distributions}",
    eprint = "2212.03878",
    archivePrefix = "arXiv",
    primaryClass = "hep-ph",
    reportNumber = "IPPP/22/79",
    doi = "10.1103/PhysRevD.108.015005",
    journal = "Phys. Rev. D",
    volume = "108",
    number = "1",
    pages = "015005",
    year = "2023"
}

@article{Carr:2020gox,
    author = "Carr, Bernard and Kohri, Kazunori and Sendouda, Yuuiti and Yokoyama, Jun'ichi",
    title = "{Constraints on primordial black holes}",
    eprint = "2002.12778",
    archivePrefix = "arXiv",
    primaryClass = "astro-ph.CO",
    reportNumber = "RESCEU-03/20; KEK-Cosmo-249; KEK-TH-2199; IPMU20-0024",
    doi = "10.1088/1361-6633/ac1e31",
    journal = "Rept. Prog. Phys.",
    volume = "84",
    number = "11",
    pages = "116902",
    year = "2021"
}

@article{Carr:1975qj,
    author = "Carr, Bernard J.",
    title = "{The Primordial black hole mass spectrum}",
    doi = "10.1086/153853",
    journal = "Astrophys. J.",
    volume = "201",
    pages = "1--19",
    year = "1975"
}

@article{Inomata:2020lmk,
    author = "Inomata, Keisuke and Kawasaki, Masahiro and Mukaida, Kyohei and Terada, Takahiro and Yanagida, Tsutomu T.",
    title = "{Gravitational Wave Production right after a Primordial Black Hole Evaporation}",
    eprint = "2003.10455",
    archivePrefix = "arXiv",
    primaryClass = "astro-ph.CO",
    reportNumber = "IPMU 20-0029, DESY 20-042, DESY-20-042, CTPU-PTC-20-05",
    doi = "10.1103/PhysRevD.101.123533",
    journal = "Phys. Rev. D",
    volume = "101",
    number = "12",
    pages = "123533",
    year = "2020"
}

@article{Papanikolaou:2020qtd,
    author = "Papanikolaou, Theodoros and Vennin, Vincent and Langlois, David",
    title = "{Gravitational waves from a universe filled with primordial black holes}",
    eprint = "2010.11573",
    archivePrefix = "arXiv",
    primaryClass = "astro-ph.CO",
    doi = "10.1088/1475-7516/2021/03/053",
    journal = "JCAP",
    volume = "03",
    pages = "053",
    year = "2021"
}

@article{Domenech:2020ssp,
    author = "Dom{\`e}nech, Guillem and Lin, Chunshan and Sasaki, Misao",
    title = "{Gravitational wave constraints on the primordial black hole dominated early universe}",
    eprint = "2012.08151",
    archivePrefix = "arXiv",
    primaryClass = "gr-qc",
    reportNumber = "YITP-20-156",
    doi = "10.1088/1475-7516/2021/11/E01",
    journal = "JCAP",
    volume = "04",
    pages = "062",
    year = "2021",
    note = "[Erratum: JCAP 11, E01 (2021)]"
}

@article{Planck:2018jri,
    author = "Akrami, Y. and others",
    collaboration = "Planck",
    title = "{Planck 2018 results. X. Constraints on inflation}",
    eprint = "1807.06211",
    archivePrefix = "arXiv",
    primaryClass = "astro-ph.CO",
    doi = "10.1051/0004-6361/201833887",
    journal = "Astron. Astrophys.",
    volume = "641",
    pages = "A10",
    year = "2020"
}

@article{BICEP:2021xfz,
    author = "Ade, P. A. R. and others",
    collaboration = "BICEP, Keck",
    title = "{Improved Constraints on Primordial Gravitational Waves using Planck, WMAP, and BICEP/Keck Observations through the 2018 Observing Season}",
    eprint = "2110.00483",
    archivePrefix = "arXiv",
    primaryClass = "astro-ph.CO",
    doi = "10.1103/PhysRevLett.127.151301",
    journal = "Phys. Rev. Lett.",
    volume = "127",
    number = "15",
    pages = "151301",
    year = "2021"
}

@article{Kohri:1999ex,
    author = "Kohri, Kazunori and Yokoyama, Jun'ichi",
    title = "{Primordial black holes and primordial nucleosynthesis. 1. Effects of hadron injection from low mass holes}",
    eprint = "astro-ph/9908160",
    archivePrefix = "arXiv",
    reportNumber = "OU-TAP-101",
    doi = "10.1103/PhysRevD.61.023501",
    journal = "Phys. Rev. D",
    volume = "61",
    pages = "023501",
    year = "2000"
}

@article{Saikawa:2018rcs,
    author = "Saikawa, Ken'ichi and Shirai, Satoshi",
    title = "{Primordial gravitational waves, precisely: The role of thermodynamics in the Standard Model}",
    eprint = "1803.01038",
    archivePrefix = "arXiv",
    primaryClass = "hep-ph",
    reportNumber = "IPMU18-0037, MPP-2018-19",
    doi = "10.1088/1475-7516/2018/05/035",
    journal = "JCAP",
    volume = "05",
    pages = "035",
    year = "2018"
}

@article{Auffinger:2020afu,
    author = "Auffinger, J{\'e}r{\'e}my and Masina, Isabella and Orlando, Giorgio",
    title = "{Bounds on warm dark matter from Schwarzschild primordial black holes}",
    eprint = "2012.09867",
    archivePrefix = "arXiv",
    primaryClass = "hep-ph",
    doi = "10.1140/epjp/s13360-021-01247-9",
    journal = "Eur. Phys. J. Plus",
    volume = "136",
    number = "2",
    pages = "261",
    year = "2021"
}

@article{Masina:2020xhk,
    author = "Masina, Isabella",
    title = "{Dark matter and dark radiation from evaporating primordial black holes}",
    eprint = "2004.04740",
    archivePrefix = "arXiv",
    primaryClass = "hep-ph",
    doi = "10.1140/epjp/s13360-020-00564-9",
    journal = "Eur. Phys. J. Plus",
    volume = "135",
    number = "7",
    pages = "552",
    year = "2020"
}

@article{Planck:2018vyg,
    author = "Aghanim, N. and others",
    collaboration = "Planck",
    title = "{Planck 2018 results. VI. Cosmological parameters}",
    eprint = "1807.06209",
    archivePrefix = "arXiv",
    primaryClass = "astro-ph.CO",
    doi = "10.1051/0004-6361/201833910",
    journal = "Astron. Astrophys.",
    volume = "641",
    pages = "A6",
    year = "2020",
    note = "[Erratum: Astron.Astrophys. 652, C4 (2021)]"
}

@article{SimonsObservatory:2018koc,
    author = "Ade, Peter and others",
    collaboration = "Simons Observatory",
    title = "{The Simons Observatory: Science goals and forecasts}",
    eprint = "1808.07445",
    archivePrefix = "arXiv",
    primaryClass = "astro-ph.CO",
    doi = "10.1088/1475-7516/2019/02/056",
    journal = "JCAP",
    volume = "02",
    pages = "056",
    year = "2019"
}

@article{Depta:2020mhj,
    author = "Depta, Paul Frederik and Hufnagel, Marco and Schmidt-Hoberg, Kai",
    title = "{ACROPOLIS: A generiC fRamework fOr Photodisintegration Of LIght elementS}",
    eprint = "2011.06518",
    archivePrefix = "arXiv",
    primaryClass = "hep-ph",
    reportNumber = "DESY-20-161, DESY 20-161, ULB-TH/20-16",
    doi = "10.1088/1475-7516/2021/03/061",
    journal = "JCAP",
    volume = "03",
    pages = "061",
    year = "2021"
}

@article{Depta:2020zbh,
    author = "Depta, Paul Frederik and Hufnagel, Marco and Schmidt-Hoberg, Kai",
    title = "{Updated BBN constraints on electromagnetic decays of MeV-scale particles}",
    eprint = "2011.06519",
    archivePrefix = "arXiv",
    primaryClass = "hep-ph",
    reportNumber = "DESY-20-160, DESY 20-160, ULB-TH/20-15",
    doi = "10.1088/1475-7516/2021/04/011",
    journal = "JCAP",
    volume = "04",
    pages = "011",
    year = "2021"
}

@article{Hufnagel:2018bjp,
    author = "Hufnagel, Marco and Schmidt-Hoberg, Kai and Wild, Sebastian",
    title = "{BBN constraints on MeV-scale dark sectors. Part II. Electromagnetic decays}",
    eprint = "1808.09324",
    archivePrefix = "arXiv",
    primaryClass = "hep-ph",
    reportNumber = "DESY 18-133, DESY-18-133",
    doi = "10.1088/1475-7516/2018/11/032",
    journal = "JCAP",
    volume = "11",
    pages = "032",
    year = "2018"
}

@article{Kawasaki:2017bqm,
    author = "Kawasaki, Masahiro and Kohri, Kazunori and Moroi, Takeo and Takaesu, Yoshitaro",
    title = "{Revisiting Big-Bang Nucleosynthesis Constraints on Long-Lived Decaying Particles}",
    eprint = "1709.01211",
    archivePrefix = "arXiv",
    primaryClass = "hep-ph",
    reportNumber = "KEK-COSMO-211, IPMU17-0117, UT-17-29, KEK-Cosmo-211, KEK-TH-1998",
    doi = "10.1103/PhysRevD.97.023502",
    journal = "Phys. Rev. D",
    volume = "97",
    number = "2",
    pages = "023502",
    year = "2018"
}

@article{Chluba:2013wsa,
    author = "Chluba, Jens",
    title = "{Distinguishing different scenarios of early energy release with spectral distortions of the cosmic microwave background}",
    eprint = "1304.6121",
    archivePrefix = "arXiv",
    primaryClass = "astro-ph.CO",
    doi = "10.1093/mnras/stt1733",
    journal = "Mon. Not. Roy. Astron. Soc.",
    volume = "436",
    pages = "2232--2243",
    year = "2013"
}

@article{Balazs:2022tjl,
    author = "Bal{\'a}zs, Csaba and others",
    title = "{Cosmological constraints on decaying axion-like particles: a global analysis}",
    eprint = "2205.13549",
    archivePrefix = "arXiv",
    primaryClass = "astro-ph.CO",
    reportNumber = "gambit-physics-2022, KCL-PH-TH/2022-23, TTP22-034",
    doi = "10.1088/1475-7516/2022/12/027",
    journal = "JCAP",
    volume = "12",
    pages = "027",
    year = "2022"
}

@article{Bolliet:2020ofj,
    author = "Bolliet, Boris and Chluba, Jens and Battye, Richard",
    title = "{Spectral distortion constraints on photon injection from low-mass decaying particles}",
    eprint = "2012.07292",
    archivePrefix = "arXiv",
    primaryClass = "astro-ph.CO",
    doi = "10.1093/mnras/stab1997",
    journal = "Mon. Not. Roy. Astron. Soc.",
    volume = "507",
    number = "3",
    pages = "3148--3178",
    year = "2021"
}

@article{Slatyer:2016qyl,
    author = "Slatyer, Tracy R. and Wu, Chih-Liang",
    title = "{General Constraints on Dark Matter Decay from the Cosmic Microwave Background}",
    eprint = "1610.06933",
    archivePrefix = "arXiv",
    primaryClass = "astro-ph.CO",
    reportNumber = "MIT-CTP-4842",
    doi = "10.1103/PhysRevD.95.023010",
    journal = "Phys. Rev. D",
    volume = "95",
    number = "2",
    pages = "023010",
    year = "2017"
}

@article{Poulin:2016anj,
    author = "Poulin, Vivian and Lesgourgues, Julien and Serpico, Pasquale D.",
    title = "{Cosmological constraints on exotic injection of electromagnetic energy}",
    eprint = "1610.10051",
    archivePrefix = "arXiv",
    primaryClass = "astro-ph.CO",
    doi = "10.1088/1475-7516/2017/03/043",
    journal = "JCAP",
    volume = "03",
    pages = "043",
    year = "2017"
}

@article{Cang:2020exa,
    author = "Cang, Junsong and Gao, Yu and Ma, Yin-Zhe",
    title = "{Probing dark matter with future CMB measurements}",
    eprint = "2002.03380",
    archivePrefix = "arXiv",
    primaryClass = "astro-ph.CO",
    doi = "10.1103/PhysRevD.102.103005",
    journal = "Phys. Rev. D",
    volume = "102",
    number = "10",
    pages = "103005",
    year = "2020"
}

@article{Baldes:2020nuv,
    author = "Baldes, Iason and Decant, Quentin and Hooper, Deanna C. and Lopez-Honorez, Laura",
    title = "{Non-Cold Dark Matter from Primordial Black Hole Evaporation}",
    eprint = "2004.14773",
    archivePrefix = "arXiv",
    primaryClass = "astro-ph.CO",
    reportNumber = "ULB-TH/20-05",
    doi = "10.1088/1475-7516/2020/08/045",
    journal = "JCAP",
    volume = "08",
    pages = "045",
    year = "2020"
}

@article{Schiavone:2021imu,
    author = "Schiavone, Francesco and Montanino, Daniele and Mirizzi, Alessandro and Capozzi, Francesco",
    title = "{Axion-like particles from primordial black holes shining through the Universe}",
    eprint = "2107.03420",
    archivePrefix = "arXiv",
    primaryClass = "hep-ph",
    doi = "10.1088/1475-7516/2021/08/063",
    journal = "JCAP",
    volume = "08",
    pages = "063",
    year = "2021"
}

@article{Langhoff:2022bij,
    author = "Langhoff, Kevin and Outmezguine, Nadav Joseph and Rodd, Nicholas L.",
    title = "{Irreducible Axion Background}",
    eprint = "2209.06216",
    archivePrefix = "arXiv",
    primaryClass = "hep-ph",
    reportNumber = "CERN-TH-2022-148",
    doi = "10.1103/PhysRevLett.129.241101",
    journal = "Phys. Rev. Lett.",
    volume = "129",
    number = "24",
    pages = "241101",
    year = "2022"
}

@article{Ayala:2014pea,
    author = "Ayala, Adrian and Dom{\'\i}nguez, Inma and Giannotti, Maurizio and Mirizzi, Alessandro and Straniero, Oscar",
    title = "{Revisiting the bound on axion-photon coupling from Globular Clusters}",
    eprint = "1406.6053",
    archivePrefix = "arXiv",
    primaryClass = "astro-ph.SR",
    doi = "10.1103/PhysRevLett.113.191302",
    journal = "Phys. Rev. Lett.",
    volume = "113",
    number = "19",
    pages = "191302",
    year = "2014"
}

@article{Dolan:2022kul,
    author = "Dolan, Matthew J. and Hiskens, Frederick J. and Volkas, Raymond R.",
    title = "{Advancing globular cluster constraints on the axion-photon coupling}",
    eprint = "2207.03102",
    archivePrefix = "arXiv",
    primaryClass = "hep-ph",
    doi = "10.1088/1475-7516/2022/10/096",
    journal = "JCAP",
    volume = "10",
    pages = "096",
    year = "2022"
}

@article{Payez:2014xsa,
    author = "Payez, Alexandre and Evoli, Carmelo and Fischer, Tobias and Giannotti, Maurizio and Mirizzi, Alessandro and Ringwald, Andreas",
    title = "{Revisiting the SN1987A gamma-ray limit on ultralight axion-like particles}",
    eprint = "1410.3747",
    archivePrefix = "arXiv",
    primaryClass = "astro-ph.HE",
    reportNumber = "DESY-14-164",
    doi = "10.1088/1475-7516/2015/02/006",
    journal = "JCAP",
    volume = "02",
    pages = "006",
    year = "2015"
}

@article{CAST:2017uph,
    author = "Anastassopoulos, V. and others",
    collaboration = "CAST",
    title = "{New CAST Limit on the Axion-Photon Interaction}",
    eprint = "1705.02290",
    archivePrefix = "arXiv",
    primaryClass = "hep-ex",
    doi = "10.1038/nphys4109",
    journal = "Nature Phys.",
    volume = "13",
    pages = "584--590",
    year = "2017"
}

@article{Moretti:2008hs,
    author = "Moretti, A. and others",
    editor = "Rodriguez, Jerome and Ferrando, Phillippe",
    title = "{A new measurement of the cosmic X-ray background}",
    eprint = "0811.1444",
    archivePrefix = "arXiv",
    primaryClass = "astro-ph",
    doi = "10.1063/1.3149419",
    journal = "AIP Conf. Proc.",
    volume = "1126",
    number = "1",
    pages = "223--226",
    year = "2009"
}

@article{Ajello:2008xb,
    author = "Ajello, M. and others",
    title = "{Cosmic X-ray background and Earth albedo Spectra with Swift/BAT}",
    eprint = "0808.3377",
    archivePrefix = "arXiv",
    primaryClass = "astro-ph",
    doi = "10.1086/592595",
    journal = "Astrophys. J.",
    volume = "689",
    pages = "666",
    year = "2008"
}

@article{Cirelli:2010xx,
    author = "Cirelli, Marco and Corcella, Gennaro and Hektor, Andi and Hutsi, Gert and Kadastik, Mario and Panci, Paolo and Raidal, Martti and Sala, Filippo and Strumia, Alessandro",
    title = "{PPPC 4 DM ID: A Poor Particle Physicist Cookbook for Dark Matter Indirect Detection}",
    eprint = "1012.4515",
    archivePrefix = "arXiv",
    primaryClass = "hep-ph",
    reportNumber = "CERN-PH-TH-2010-057, SACLAY-T10-025, IFUP-TH-2010-44",
    doi = "10.1088/1475-7516/2012/10/E01",
    journal = "JCAP",
    volume = "03",
    pages = "051",
    year = "2011",
    note = "[Erratum: JCAP 10, E01 (2012)]"
}

\clearpage
\onecolumngrid
\begin{center}
   \textbf{\large SUPPLEMENTAL MATERIAL \\[.2cm] ``Hawking Axions''}\\[.2cm]
  \vspace{0.05in}
  {Abdelaziz Hussein, Gordan Krnjaic, and Sarunas Verner}
\end{center}

\setcounter{equation}{0}
\setcounter{figure}{0}
\setcounter{table}{0}
\setcounter{section}{0}
\setcounter{subsection}{0}
\setcounter{page}{1}

\setcounter{secnumdepth}{2}

\renewcommand{\theequation}{S\arabic{equation}}
\renewcommand{\thefigure}{S\arabic{figure}}
\renewcommand{\thetable}{S\arabic{table}}

\renewcommand{\thesection}{\arabic{section}}
\renewcommand{\thesubsection}{\Alph{subsection}}

\onecolumngrid
In this Supplemental Material we collect the technical details behind the results of the paper. Section~\ref{sec:background} describes the PBH background and the resulting ALP distribution, Section~\ref{sec:decay} the treatment of ALP decays, Section~\ref{sec:constraints} how each constraint is implemented, and Section~\ref{sec:xray} the X-ray bound. 

\section{PBH background and the ALP distribution}
\label{sec:background}

\subsection{Assumptions}
Throughout we work in natural units with $G=\mpl^{-2}$ and normalize the scale factor so that $a_0=1$. For the late-time cosmology we adopt the Planck 2018 parameters~\cite{Planck:2018vyg}. We take all PBHs to have the same initial mass $M_i$ and to be nonrotating, and we ignore accretion and mergers. This is a reasonable starting point, since PBHs that form from rare peaks during radiation domination are expected to be born with small spins~\cite{DeLuca:2019buf,Mirbabayi:2019uph}. A nonzero spin, or a spread in the initial masses, would modify both the evaporation rate and the spectrum of the emitted particles~\cite{Page:1976ki,Arbey:2021ysg,Cheek:2022mmy}, and we do not consider these cases here. The SM particles produced by evaporation thermalize by the time the PBHs disappear~\cite{Shallue:2024hqe}, whereas the ALPs interact only through their decay into photons. Our results therefore apply only where plasma effects can neither thermalize nor reabsorb the ALPs.

We also treat the production and the decay of the ALPs as two separate stages. For this to be valid, decays during evaporation have to be negligible, which requires $\Gamma_a\tau_{\rm BH}\ll1$. In practice this condition is conservative, because the ALPs are relativistic when they are emitted and time dilation only slows down their decay. For the graybody factors we use the tabulated massless spin-zero results, which limits us to $\ma/T_{\rm BH}^{\,i}\lesssim0.1$ with $T_{\rm BH}^{\,i}\equiv T_{\rm BH}(M_i)$. This is easily satisfied over most of our parameter space. For instance, for $\ma\leq100\,$MeV the ratio never exceeds $5\times10^{-6}$ in our mass window, while for $M_i=10^8\,$g and $\ma=10\,$TeV it reaches $0.094$, close to the edge of the allowed range. Apart from the graybody factors, the ALP mass is kept everywhere in the calculation, including the dispersion relation, the production threshold, and the decay kinematics. We note that ALPs heavier than $T_{\rm BH}^{\,i}$ are not forbidden, since they are emitted at late times once the PBH has heated up. Describing this regime would however require massive graybody factors, and we leave it for future work.

\subsection{Mass window and lifetime}
It is convenient to write the mass-loss rate of Eq.~\eqref{eq:massloss-main} as $dM/dt=-\kappa/M^2$. For the SM we have $\kappa={\cal G}g_{\star,H}\mpl^4/(30720\pi)\simeq8\times10^{26}\,{\rm g^3\,s^{-1}}$. Since all SM species are already emitted at the beginning of evaporation (even for $M_i=5\times10^8\,$g the initial temperature is $T_{\rm BH}^{\,i}\simeq2\times10^4\,$GeV), $\kappa$ is to a good approximation constant. The PBH mass then decreases as $M(t)=M_i(1-t/\tau_{\rm BH})^{1/3}$, with a lifetime
\begin{equation}
 \tau_{\rm BH}=\frac{M_i^3}{3\kappa}\simeq0.42\,{\rm s}\left(\frac{M_i}{10^9\,{\rm g}}\right)^3 .
 \label{eq:S-lifetime}
\end{equation}
A light real scalar takes away about $1.7\%$ of the Hawking power~\cite{Page:1976df,MacGibbon:1990zk,Shallue:2024hqe}, so its effect on $\kappa$ can be safely neglected.

To determine the allowed range of $M_i$, we assume that the PBHs form during radiation domination with a mass equal to a fraction $\gamma\simeq0.2$ of the horizon mass~\cite{Carr:1975qj,Carr:2020gox},
\begin{equation}
 M_i=\frac{\gamma}{2GH_f}\simeq2.7\times10^4\,{\rm g}\left(\frac{\gamma}{0.2}\right)\left(\frac{10^9\,{\rm GeV}}{H_f}\right).
 \label{eq:S-formation}
\end{equation}
Since formation has to happen after inflation, the Planck and BICEP/Keck bound $H_f\lesssim5\times10^{13}\,$GeV~\cite{Planck:2018jri,BICEP:2021xfz} gives $M_i\gtrsim0.5\,$g, and we take $M_i=1\,$g as the lower end of the window. The upper end comes from BBN, which needs a reheating temperature of at least a few MeV~\cite{Kohri:1999ex,deSalas:2015glj,Hasegawa:2019jsa}. Using Eq.~\eqref{eq:S-TRH} below, this translates into $M_i\lesssim5\times10^8\,$g~\cite{Shallue:2024hqe}.

\subsection{Attractor and reheating}
If the PBHs form with an initial energy fraction $\beta$, they come to dominate the energy density before they evaporate provided that $\beta\gtrsim\beta_c\simeq(t_f/\tau_{\rm BH})^{1/2}\simeq5\times10^{-6}\,({\rm g}/M_i)$, where we used $t_f\simeq GM_i/\gamma$. When $\beta\gg\beta_c$, the evolution near evaporation no longer depends on $\beta$, and neither do the reheating temperature and the normalized ALP distribution~\cite{Barrow:1990he,Hooper:2019gtx,Shallue:2024hqe}. Gravitational wave constraints on fluctuations in the PBH number density still allow this regime throughout our mass window~\cite{Inomata:2020lmk,Papanikolaou:2020qtd,Domenech:2020ssp}. In practice, we therefore solve Eq.~\eqref{eq:background-main} starting from $\rho_R=0$, as described in the paper. Right after evaporation the solution gives
\begin{equation}
 \rho_R(t_{\rm ev})\simeq\frac{0.88}{6\pi G\tau_{\rm BH}^2},
 \qquad
 \widehat n_{\rm BH}\simeq\frac{1.09\,a_{\rm RH}^3}{6\pi G\tau_{\rm BH}^2M_i},
 \label{eq:S-attractor}
\end{equation}
where $a_{\rm RH}\equiv a(t_{\rm ev})$ and $\widehat n_{\rm BH}=a^3n_{\rm BH}$ is constant until the PBHs evaporate. We have checked that adding primordial radiation with $10\%$ of the PBH density to the initial conditions changes these coefficients by less than $0.02\%$.

Not all of the radiation at $t_{\rm ev}$ is thermal. A fraction $\epsilon_{\rm dec}\simeq\epsilon_a$ of it is carried by the decoupled ALPs, and a much smaller amount by gravitons. We therefore define the reheating temperature using only the SM part, $(1-\epsilon_{\rm dec})\rho_R(t_{\rm ev})=\pi^2g_\star(T_{\rm RH})T_{\rm RH}^4/30$, which leads to
\begin{equation}
 T_{\rm RH}\simeq2.7\times10^{10}\,{\rm GeV}\left[\frac{106.75}{g_\star(T_{\rm RH})}\right]^{1/4}\left(\frac{1\,{\rm g}}{M_i}\right)^{3/2}.
 \label{eq:S-TRH}
\end{equation}
The factor $(1-\epsilon_{\rm dec})^{1/4}$ changes $T_{\rm RH}$ by only $0.4\%$. Because $g_\star$ depends on $T_{\rm RH}$, we solve Eq.~\eqref{eq:S-TRH} iteratively, using the SM equation of state of Ref.~\cite{Saikawa:2018rcs}. If the ALP decays do not produce a significant amount of entropy, entropy conservation after reheating gives
\begin{equation}
 a_{\rm RH}=\frac{T_{\gamma0}}{T_{\rm RH}}\left[\frac{g_{\star s,0}}{g_{\star s}(T_{\rm RH})}\right]^{1/3}\simeq2.9\times10^{-24}\,\eta\left(\frac{M_i}{1\,{\rm g}}\right)^{3/2},
 \qquad
 \eta\equiv\left[\frac{106.75}{g_{\star s}(T_{\rm RH})}\right]^{1/3}\left[\frac{g_\star(T_{\rm RH})}{106.75}\right]^{1/4},
 \label{eq:S-aRH}
\end{equation}
where $T_{\gamma0}=2.7255\,$K and $g_{\star s,0}\simeq3.91$ does not include the Hawking relics. For $M_i\lesssim10^5\,$g we find $\eta\simeq1$. For heavier PBHs $\eta$ grows slowly and reaches about $1.2$ at $5\times10^8\,$g, since fewer SM species are left to transfer their entropy to the photons. Together, Eqs.~\eqref{eq:S-attractor} and~\eqref{eq:S-aRH} fix the overall normalization of $f_a$ in Eq.~\eqref{eq:phase-space-main}.

\subsection{The ALP distribution}
To compute $f_a$ from Eq.~\eqref{eq:phase-space-main}, we use the primary spin-zero spectrum of \textsc{BlackHawk}~v2.3~\cite{Arbey:2019mbc,Arbey:2021mbl}, evaluate the graybody factor at the instantaneous mass $M(t)$, and take $a(t)$ from the numerical background solution. In this way each ALP is redshifted from the moment it is emitted, which would not be the case if we first integrated the spectrum over the PBH lifetime.

The shape of $f_a$ can be understood from a simple estimate. If we ignore the expansion and use the geometric-optics graybody factor, a change of variables to $x=p/T_{\rm BH}(M)$ gives
\begin{equation}
 \frac{dn_a}{dp}\simeq\frac{27\,n_{\rm BH}}{2\pi(8\pi)^5\kappa G^3p^3}\int_0^{p/T_{\rm BH}^{\,i}}dx\,\frac{x^4}{e^x-1}.
 \label{eq:S-static}
\end{equation}
For $p\ll T_{\rm BH}^{\,i}$ the integral grows as $(p/T_{\rm BH}^{\,i})^4$ and $dn_a/dp\propto p$, while for $p\gg T_{\rm BH}^{\,i}$ it approaches a constant and $dn_a/dp\propto p^{-3}$. The high-momentum tail is produced at the very end of evaporation, when the temperature of the PBH rises quickly (half of the PBH mass is lost during the last eighth of its lifetime). With the full graybody factor both power laws survive, and only their coefficients change. The expansion of the universe moves the ALPs emitted early to smaller comoving momenta, but it barely affects the tail, because those ALPs are emitted shortly before the PBHs disappear.

In the PBH-dominated attractor and for $\ma\ll T_{\rm BH}^{\,i}$, the distributions for different PBH masses are related by a simple rescaling~\cite{Shallue:2024hqe}. In particular, the comoving number density and the mean momentum today scale as
\begin{equation}
 \widehat n_a\propto\eta^3M_i^{-1/2},
 \qquad
 \langle q\rangle\simeq0.14\,{\rm eV}\,\eta\left(\frac{M_i}{1\,{\rm g}}\right)^{1/2}.
 \label{eq:S-scaling}
\end{equation}
The mean momentum corresponds to $\langle p\rangle\simeq4.6\,T_{\rm BH}^{\,i}$ at reheating. For comparison, a scalar that decoupled from the SM plasma above the electroweak scale would have $\langle q\rangle\sim10^{-4}\,$eV today, so the Hawking population is much hotter, and also much less abundant. If the ALP is stable and nonrelativistic today, its abundance is
\begin{equation}
 \Omega_ah^2\simeq2.0\times10^{-6}\,\eta^3\left(\frac{\ma}{\rm eV}\right)\left(\frac{1\,{\rm g}}{M_i}\right)^{1/2},
 \label{eq:S-omega}
\end{equation}
while if it is still relativistic after electron--positron annihilation, it contributes
\begin{equation}
 \Delta N_{\rm eff}=\frac{8}{7}\left(\frac{11}{4}\right)^{4/3}\frac{\rho_a}{\rho_\gamma}\simeq0.051\,\eta^4
 \label{eq:S-neff}
\end{equation}
to the effective number of neutrinos~\cite{Hooper:2019gtx,Auffinger:2020afu,Masina:2020xhk,Arbey:2021ysg,Shallue:2024hqe}. They can be compared with the Planck bound $\Delta N_{\rm eff}\lesssim0.3$ at $95\%$ C.L.~\cite{Planck:2018vyg} and with the Simons Observatory forecast $\sigma(N_{\rm eff})\simeq0.05$--$0.07$~\cite{SimonsObservatory:2018koc}. For a decaying ALP, however, $N_{\rm eff}$ depends on both the surviving ALPs and the heating of the photons.

We tested the numerical spectra in several ways. At high energies we compared them with the geometric optics result, and we checked that the spectra for different PBH masses follow Eq.~\eqref{eq:S-scaling}. We also verified energy conservation, $\int dq\,q\,f_a=a_{\rm RH}^4\rho_a(t_{\rm ev})$, which has to reproduce the ALP fraction $\epsilon_a$, and that doubling the time and momentum resolution leaves the moments of $f_a$ unchanged.

\section{Decays and the photon spectrum}
\label{sec:decay}
For the survival probability in Eq.~\eqref{eq:survival-main} we write the exponent in terms of $x=\ln(1+z)$,
\begin{equation}
 {\cal D}_a(q,z)=\Gamma_a\ma\int_{\ln(1+z)}^{\ln(1+z_{\rm RH})}\frac{dx'}{H(x')\sqrt{\ma^2+q^2e^{2x'}}},
 \qquad S_a=e^{-{\cal D}_a},
 \label{eq:S-depth}
\end{equation}
and evaluate the integral for each momentum mode, starting from reheating. The expansion rate includes radiation, with the temperature dependence of $g_\star$ and $g_{\star s}$ in the SM, as well as matter and dark energy. As a check, we compared the result with the analytic expression that can be derived for a universe that contains only radiation.

The box kernel in Eq.~\eqref{eq:box-main} is normalized to two photons per decay, $\int dE_\gamma\,K_\gamma=2$, and its first moment gives back the ALP energy, $\int dE_\gamma\,E_\gamma K_\gamma=E_a$. Since $E_a/\gamma_a=\ma$, the larger energy of the photons from a boosted decay exactly compensates the time dilation of the decay rate, and the injection rate in Eq.~\eqref{eq:inputs-main2} simplifies to
\begin{equation}
 \dot\rho_{\rm inj}(z)=\ma\Gamma_a(1+z)^3\int dq\,f_a(q)\,S_a(q,z).
 \label{eq:S-injection}
\end{equation}
The ALP momentum therefore enters only through $S_a$. Numerically, we compute the lower edge of the box as $E_-=\ma^2/[2(E_a+p)]$ to avoid a cancellation when $E_a\simeq p$.

\section{Constraints}
\label{sec:constraints}

\subsection{Existing bounds}
We take the existing limits on $\gagg$ from the \textsc{AxionLimits} repository~\cite{AxionLimits}, but keep only those that do not depend on the cosmological ALP abundance. These are the helioscope, stellar evolution, solar, and supernova bounds, including the limits from the conversion of ALPs produced in these sources~\cite{Ayala:2014pea,Dolan:2022kul,Payez:2014xsa,CAST:2017uph}. We do not include cosmological bounds that assume thermal production, freeze-in, or a fixed relic abundance, and neither haloscope nor dark matter decay limits, which would have to be recast for the Hawking population. The gray regions in Figs.~\ref{fig:constraints-heavy} and~\ref{fig:constraints-light} show the union of the remaining limits. Note that these are not all derived at the same confidence level.

\subsection{ALP domination}
We exclude every point where the ALP density $\rho_a$ of Eq.~\eqref{eq:inputs-main1} exceeds the standard background density, $\rho_{\rm bg}=\rho_{\rm rad}+\Omega_m\rho_c(1+z)^3+\Omega_\Lambda\rho_c$, at some redshift between reheating and today. Relativistic ALPs never dominate, since they make up only about $1\%$ of the radiation density. Once they become nonrelativistic, however, their density grows relative to radiation. Stable ALPs therefore dominate if $\ma\widehat n_a\gtrsim\Omega_m\rho_c$, that is, if their density today would exceed the observed matter density. For shorter lifetimes the ALPs have to dominate before they decay. During radiation domination the ratio $\rho_a/\rho_{\rm rad}$ grows as $\ma\widehat n_a t^{1/2}$, so the minimum lifetime for domination scales as $(\ma\widehat n_a)^{-2}$. With $\tau_a=64\pi/(\gagg^2\ma^3)$, the boundary then scales as $\gagg\propto\widehat n_a\ma^{-1/2}$.

\subsection{BBN}
For the BBN constraint we tabulate the physical photon source $(1+z)^3\widehat Q_\gamma(E_\gamma,z)$ on logarithmic grids in redshift and photon energy, convert redshift to time with the same $H(z)$ as above, and use it as input for \textsc{ACROPOLIS}~\cite{Hufnagel:2018bjp,Depta:2020zbh,Depta:2020mhj}. The code solves the electromagnetic cascade without assuming a universal spectrum, and then evolves the light-element abundances after standard BBN. Pair production on the thermal photons suppresses the cascade spectrum above $E_{\rm C}\simeq m_e^2/(22T)$~\cite{Kawasaki:2017bqm}, so deuterium, with a binding energy of $2.22\,$MeV, can only be destroyed once the universe has cooled enough for $E_{\rm C}$ to exceed this value. For boosted ALPs the photons can lie above threshold even when $\ma/2$ is below it. We start from the standard BBN abundances of the reference \textsc{ACROPOLIS} analysis and exclude a point if D/H or $Y_p$ falls outside its $95\%$ range, or if $^3$He/D is above its upper limit~\cite{Depta:2020zbh,Depta:2020mhj}. Because of the unresolved lithium problem, we do not use $^7$Li. To test the input, we checked that each tabulated source reproduces the photon number and energy moments, $2\int dq\,f_aS_a\Gamma_a/\gamma_a$ and $\dot\rho_{\rm inj}/(1+z)^3$, and that the standard abundances are recovered when the source is switched off. We also refined the grids until the contour stopped changing. Since we use standard initial abundances and a fixed background, the calculation is valid only if the ALPs neither dominate nor release a significant amount of entropy before photodisintegration becomes efficient.

\subsection{Spectral distortions}
For the spectral distortions we assume that the injected energy is converted into heat right away. The fractional energy release per unit redshift is then ${\cal E}(z)=\dot\rho_{\rm inj}/[\rho_\gamma(1+z)H]$, and the distortion parameters are
\begin{equation}
 \mu=1.401\int_{z_{\rm rec}}^{z_{\rm RH}}dz\,{\cal J}_\mu{\cal E},
 \qquad
 y=\frac14\int_{z_{\rm rec}}^{z_{\rm RH}}dz\,{\cal J}_y{\cal E},
 \label{eq:S-mu-y}
\end{equation}
with $z_{\rm rec}=1100$. For the visibility functions we use~\cite{Chluba:2011hw,Chluba:2013wsa}
\begin{equation}
 {\cal J}_{\rm bb}=e^{-[(1+z)/(1+z_{\rm dc})]^{5/2}},
 \qquad
 {\cal J}_y=\frac{{\cal J}_{\rm bb}}{1+[(1+z)/(1+z_{\mu y})]^{2.58}},
 \qquad
 {\cal J}_\mu={\cal J}_{\rm bb}\left\{1-e^{-[(1+z)/(5.8\times10^4)]^{1.88}}\right\},
 \label{eq:S-visibility}
\end{equation}
where $z_{\rm dc}=1.98\times10^6$ and $z_{\mu y}=6.0\times10^4$. In the numerical integration we count the decays through the change in $S_a$ between neighboring redshifts, so that $S_a$ never has to be differentiated. The FIRAS contour corresponds to $|\mu|<9.0\times10^{-5}$ and $|y|<1.5\times10^{-5}$~\cite{Fixsen:1996nj}, applied separately and not through a fit to the FIRAS spectrum. For the PIXIE-like projection we use instead the $5\sigma$ values $\mu=5\times10^{-8}$ and $y=10^{-8}$ of Ref.~\cite{Kogut:2011xw}. At a fixed ALP mass this constraint excludes a band in coupling. For small couplings too few ALPs decay during the distortion era, whereas for large couplings most of them decay above $z_{\rm dc}$, where any distortion is erased. Our estimate neglects delayed energy deposition, as well as distortions beyond $\mu$ and $y$~\cite{Chluba:2013wsa,Bolliet:2020ofj}. Spectral distortion constraints on decaying ALPs were also discussed in Ref.~\cite{Balazs:2022tjl}.

\subsection{CMB anisotropies}
We denote by $\tau_{95}(m)$ the Planck $95\%$ lower limit on the lifetime of cold dark matter with mass $m$ that decays into two photons~\cite{Slatyer:2016qyl,Poulin:2016anj,Cang:2020exa}. A cold and long-lived ALP population that would make up a fraction $F_a=\ma\widehat n_a/(\Omega_{\rm DM}\rho_c)$ of the dark matter if it were stable is then excluded for $F_a\Gamma_a\tau_{95}(\ma)>1$. Since the photons from a boosted decay are spread over the box, we average the lifetime response over the photon energy at $z_*=300$, where the CMB is most sensitive to long-lived decays~\cite{Slatyer:2016qyl,Poulin:2016anj},
\begin{equation}
 {\cal W}_{95}(q)=\frac{1}{E_a}\int dE_\gamma\,E_\gamma K_\gamma\,\tau_{95}(2E_\gamma)=\int_0^1du\,\tau_{95}\!\left(2\sqrt{E_-^2+uE_ap}\right),
 \label{eq:S-W}
\end{equation}
with $p=q(1+z_*)$, and exclude a point if
\begin{equation}
 \frac{\Gamma_a\ma}{\Omega_{\rm DM}\rho_c}\int dq\,f_a(q)\,S_a(q,z_*)\,{\cal W}_{95}(q)>1 .
 \label{eq:S-CMB}
\end{equation}
For a cold population ${\cal W}_{95}\to\tau_{95}(\ma)$, and if depletion can be neglected, this condition reduces to $F_a\Gamma_a\tau_{95}(\ma)>1$. We interpolate $\log\tau_{95}$ linearly in $\log m$ and use only photon energies within the range of the published limit. The solid contour is shown only where the box average changes the response by less than $1\%$,
\begin{equation}
 \delta_{\rm box}\equiv\left|\frac{\int dq\,f_a{\cal W}_{95}}{\tau_{95}(\ma)\int dq\,f_a}-1\right|<0.01,
 \label{eq:S-dbox}
\end{equation}
where the population is long-lived at $z_*$, and where the ALPs never dominate. For shorter lifetimes the decays happen earlier, and this cannot be described by the survival probability at a single redshift. Finding where the bound weakens would require a full calculation of the energy deposition and of the recombination history, so we leave this region out. For the Simons Observatory projection we multiply $\tau_{95}$ by a constant factor $A_{\rm SO}$~\cite{Cang:2020exa,SimonsObservatory:2018koc} and keep everything else the same.

\subsection{Structure formation}
For structure formation we use the scalar mass bound $m_a^{\rm SF}(M_i)$ shown in Fig.~10 of Ref.~\cite{Shallue:2024hqe}. In that work the Hawking relics were mapped onto equivalent thermal relics and constrained with CMB, galaxy clustering, and weak lensing data. We interpolate the boundary in $\log M_i$ and $\log\ma$, and exclude $\ma>m_a^{\rm SF}(M_i)$ only when the decays have a small effect,
\begin{equation}
 \Delta_{\rm stab}\equiv\max\left[1-\frac{\widehat n_a^{\rm surv}}{\widehat n_a},\left|\frac{\langle v\rangle_{\rm surv}}{\langle v\rangle_0}-1\right|\right]<10^{-2}.
 \label{eq:S-stab}
\end{equation}
Here $\widehat n_a^{\rm surv}=\int dq\,f_aS_a(q,0)$, and $\langle v\rangle$ is the mean speed today, computed with and without the survival probability. The second condition is needed because fast modes live longer due to time dilation, so the decays change not only the number of ALPs but also the shape of their distribution. Points that do not satisfy Eq.~\eqref{eq:S-stab} would have to be treated with a Boltzmann code that follows both the ALPs and their decay products.

\section{X-ray background bound}
\label{sec:xray}
Inserting the box kernel into Eq.~\eqref{eq:emissivity-main}, the emissivity at a fixed emitted energy $E_\gamma=E_0(1+z)$ becomes an integral over the tail of the momentum distribution,
\begin{equation}
 \widehat Q_\gamma(E_\gamma,z)=\frac{2\ma\Gamma_a}{1+z}\int_{q_{\min}}^\infty dq\,\frac{f_a(q)S_a(q,z)}{q\,E_a(q,z)},
 \qquad
 q_{\min}=\frac{1}{1+z}\left|E_\gamma-\frac{\ma^2}{4E_\gamma}\right|,
 \label{eq:S-tail}
\end{equation}
so we do not need to sample the decay angles. For nonrelativistic ALPs, a given observed energy $E_0$ receives contributions only from a narrow redshift interval around $1+z_d=\ma/(2E_0)$, of width $\Delta\ln(1+z)\simeq q/E_0$, and we refine the redshift grid in this region. In the absence of attenuation, integrating Eq.~\eqref{eq:flux-main} over all energies gives
\begin{equation}
 \int dE_0\,\frac{d\Phi_\gamma}{dE_0}=\frac{1}{2\pi}\int dq\,f_a(q)\left[1-S_a(q,0)\right],
 \label{eq:S-number}
\end{equation}
that is, two photons for every ALP that has decayed by today. We use this relation to check the normalization of the flux.

For the photon transmission $e^{-\tau_\gamma}$ at $z\leq1000$ we use the file \texttt{OpticalDepth.m} from \textsc{PPPC4DMID} with the \texttt{minUV} background~\cite{Cirelli:2010xx}. At higher redshifts we continue it as
\begin{equation}
 e^{-\tau_\gamma(E_0,z)}=e^{-\tau_\gamma(E_0,1000)}\exp\!\left\{-7.8\times10^{-5}\left[(1+z)^{3/2}-1001^{3/2}\right]\right\},
 \label{eq:S-opacity}
\end{equation}
and set it to zero once the exponent exceeds $200$. The upper edge of the CXB band depends on this choice. Photons produced in electromagnetic cascades are not included in the direct signal.

\begin{figure}[t]
\centering
\includegraphics[width=0.60\textwidth]{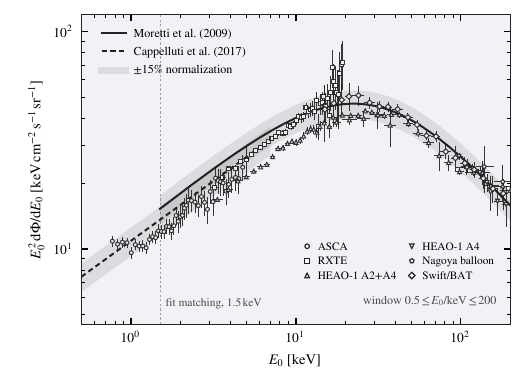}
\caption{Reference cosmic X-ray background used in this work. The dashed line shows the extragalactic power law of Cappelluti et al.~\cite{Cappelluti_2017} and the solid line the fit of Moretti et al.~\cite{Moretti:2008hs}, which we join at $1.5\,$keV (vertical line). The open symbols are the measurements compiled in Fig.~14 of Ref.~\cite{Ajello:2008xb} (ASCA, RXTE, HEAO-1 and its A4 detector, the Nagoya balloon experiment, and Swift/BAT). The bands correspond to a $\pm15\%$ change in the normalization of each fit and are shown only for illustration. The shaded region is the $0.5$--$200\,$keV window used in the analysis.}
\label{fig:cxb-reference}
\end{figure}

Below $1.5\,$keV we take the reference background from the extragalactic power law of Cappelluti et al.~\cite{Cappelluti_2017}, ${\cal B}_{\rm C}=10.91\,(E_0/{\rm keV})^{0.55}\,{\rm keV\,cm^{-2}\,s^{-1}\,sr^{-1}}$, which is corrected for Galactic absorption and does not include the Galactic foregrounds. Between $1.5$ and $200\,$keV we use the fit of Moretti et al.~\cite{Moretti:2008hs},
\begin{equation}
 {\cal B}_{\rm M}=\frac{0.109\,(E_0/{\rm keV})^2}{(E_0/29\,{\rm keV})^{1.40}+(E_0/29\,{\rm keV})^{2.88}}\,{\rm keV\,cm^{-2}\,s^{-1}\,sr^{-1}}.
 \label{eq:S-moretti}
\end{equation}
The reference background ${\cal B}_{\rm CXB}$ is ${\cal B}_{\rm C}$ below $1.5\,$keV and ${\cal B}_{\rm M}$ above. At $1.5\,$keV the two fits differ by $12\%$. We use their central values, and in Fig.~\ref{fig:cxb-reference} we compare them with the data compiled by Ajello et al.~\cite{Ajello:2008xb}.

\begin{figure*}[t]
\centering
\includegraphics[width=\textwidth]{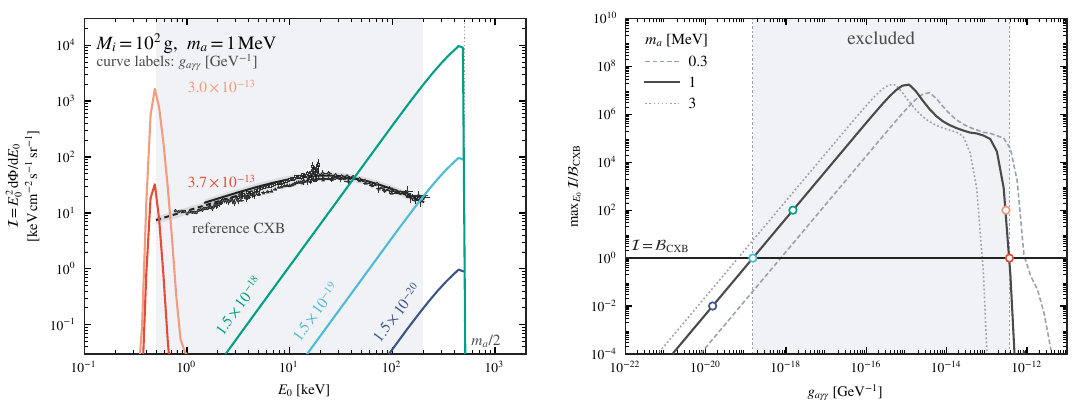}
\caption{Direct photon signal for $M_i=10^2\,$g. \emph{Left:} Attenuated spectra for $\ma=1\,$MeV and five values of the coupling (in GeV$^{-1}$), compared with the reference CXB (black). The shaded region is the analysis window. The three rising spectra correspond to long-lived ALPs. In the transparent matter-dominated era they approach ${\cal I}\propto E_0^{5/2}$, with an amplitude proportional to $\gagg^2$ and an endpoint close to $\ma/2$. For the two largest couplings most of the ALPs decay before their photons can reach us, and what remains is a narrow feature whose position depends on the decay history and the opacity. \emph{Right:} Largest ratio of the signal to the reference background in the analysis window, $\max_{E_0}{\cal I}/{\cal B}_{\rm CXB}$, as a function of the coupling for $\ma=0.3$, $1$, and $3\,$MeV. Points above one are excluded. The open circles correspond to the spectra in the left panel, and the shaded interval shows the CXB band for $\ma=1\,$MeV before the other constraints are applied.}
\label{fig:cxb-signal-examples}
\end{figure*}

Following Eq.~\eqref{eq:overshoot-main}, a point is excluded if ${\cal I}(E_0)>{\cal B}_{\rm CXB}(E_0)$ at any energy in the window $0.5\leq E_0/{\rm keV}\leq200$. The largest ratio ${\cal I}/{\cal B}_{\rm CXB}$ can occur at the edges of the window, so the energy grid includes both endpoints, as well as both sides of the $1.5\,$keV join. Figure~\ref{fig:cxb-signal-examples} shows a few examples for $M_i=10^2\,$g. For $\ma=1\,$MeV the excluded band extends from $\gagg\simeq1.5\times10^{-19}$ to $3.7\times10^{-13}\,{\rm GeV}^{-1}$. At the lower edge only a fraction $7\times10^{-8}$ of the ALPs has decayed by today and the signal grows as $\gagg^2$, while at the upper edge the decays take place when the universe is still opaque to the photons. Note that in this case Eq.~\eqref{eq:S-omega} gives $\Omega_ah^2\simeq0.20\,(\ma/{\rm MeV})$, so a long-lived population would exceed the dark matter density for $\ma\gtrsim0.6\,$MeV.

To illustrate what could be gained with a better model of the astrophysical background, we also show the reach obtained by replacing ${\cal B}_{\rm CXB}$ with $1.64\,\delta\,{\cal B}_{\rm CXB}$, for $\delta=0.1$ and $0.01$. The factor $1.64$ corresponds to a one-sided $95\%$ threshold for a Gaussian residual at a fixed energy. On the long-lived branch, where ${\cal I}\propto\gagg^2$, this improves the reach in coupling by factors of $2.5$ and $7.8$, respectively. These curves do not include correlations between energy bins, the instrumental response, or the uncertainty in modeling the astrophysical sources. Finally, we include only the isotropic extragalactic signal. A Galactic contribution would depend on the local density and velocity distribution of the surviving ALPs, which we have not computed.

\end{document}